\documentclass[pra,twocolumn]{revtex4}
\usepackage{amsmath}
\usepackage{amssymb}
\usepackage{epsfig}

\begin{document}

\title{\bf Parametric instability of oscillations of a vortex ring in a $z$-periodic 
Bose-Einstein condensate and the recurrence to starting state}
\author{Victor P. Ruban}
\email{ruban@itp.ac.ru}
\affiliation{L.D. Landau Institute for Theoretical Physics RAS, Moscow, Russia} 

\date{\today}

\begin{abstract}
The dynamics of deformations of a quantum vortex ring in a Bose-Einstein condensate with periodic
equilibrium density $\rho(z)= 1-\epsilon\cos z$ has been considered within the local induction approximation.
Parametric instabilities of the normal modes with azimuthal numbers $\pm m$ have been revealed 
at the energy integral $E$ near values $E_m^{(p)}=2m\sqrt{m^2-1}/p$, where $p$ is the
resonance order. Numerical simulations have shown that already at $\epsilon\sim 0.03$ a rapid growth 
of unstable modes with $m=2$, $p=1$ to magnitudes of order of unity is typical, which is then followed,
after a few large oscillations, by fast return to a weakly excited state. Such behavior corresponds
to an integrable Hamiltonian of the form 
$H\propto \sigma(E_2^{(1)}-E)(|b_+|^2 + |b_-|^2)  -\epsilon(b_+ b_- + b_+^* b_-^*) 
+u(|b_+|^4 +|b_-|^4) + w |b_+|^2|b_-|^2$ for two complex envelopes $b_\pm(t)$. The results have been compared
to parametric instabilities of vortex ring in condensate with density $\rho(z,r)=1-r^2-\alpha z^2$, 
which take place at $\alpha\approx 8/5$ and at $\alpha\approx 16/7$.
\end{abstract}
\maketitle

{\bf Introduction}.
The dynamics of quantum vortices in a trapped atomic Bose-Einstein condensate with spatially inhomogeneous
equilibrium density  $\rho({\mathbf r})$ differs significantly from their dynamics in a uniform
system, and the differences are not only quantitative but also qualitative (see review \cite{F2009} and
references therein). Development of experimental methods in this field makes actual new and diverse profiles  
$\rho({\mathbf r})$. Therefore, vortices in nonuniform systems continue to attract interest 
in experiment as well as in the theory \cite{SF2000,FS2001,R2001,AR2001,GP2001,A2002,RBD2002,AD2003,AD2004,SR2004,D2005,Kelvin_vaves,ring_istability,
reconn-2017,top-2017}. In the general case the problem is quite complicated, because vortices interact
with potential excitations and with non-condensate atoms. But if the condensate at zero temperature 
is in the Thomas-Fermi regime (the vortex core width $\xi$ is much smaller than a typical scale
of the inhomogeneity and the vortex size  $R_*$), then one can neglect the potential degrees 
of freedom and use the ``anelastic'' hydrodynamic approximation
\cite{SF2000,FS2001,R2001,A2002,SR2004,R2017-1,R2017-2}. If, besides that, a vortex line configuration is far 
from self-intersections, then a simple mathematical model is applicable, the local induction equation
\cite{SF2000,FS2001,R2001}
\begin{equation}
{\mathbf R}_t\big|_{\rm normal}=\frac{\Gamma \Lambda}{4\pi}\Big(\varkappa {\mathbf b}
+[\nabla\ln\rho({\mathbf R})\times {\boldsymbol \tau}]\Big),
\label{LIA}
\end{equation}
where ${\mathbf R}(\beta,t)$ is the geometric shape of the filament depending on arbitrary longitudinal parameter
$\beta$ and time  $t$, the parameter  $\Gamma=2\pi\hbar/m_{\rm atom}$ is the velocity circulation quantum,
$\Lambda=\ln(R_*/\xi)\approx$ const is a large logarithm, $\varkappa$ is a local curvature of the vortex line, 
${\mathbf b}$ is the unit binormal vector, and ${\boldsymbol \tau}$ is the unit tangent vector.
To make formulas clean, below we use dimensionless quantities, so that ${\Gamma \Lambda}/{4\pi}=1$, $R_*\sim 1$.
It is a well known fact that in the case $\rho =$ const, the local induction equation is reduced by the Hasimoto
transform \cite{Hasimoto} to the one-dimensional (1D) focusing nonlinear Schr\"odinger equation, so the vortex line 
dynamics against a uniform background is nearly integrable. For nonuniform density profiles 
investigation of this model is still in the very beginning \cite{Kelvin_vaves,ring_istability,R2016-1,R2016-2}.
Even the simplest 1D-periodic density profile
\begin{equation}
\rho(z)= 1-\epsilon\cos z 
\label{density}
\end{equation}
was not applied so far in the framework of Eq.(\ref{LIA}), although, by the way it is easily realized 
in optical traps. The purpose of this work is to fill this gap in the theory by considering the propagation
of a deformed quantum vortex ring through Bose-Einstein condensate with nonuniform density (\ref{density}). 
By theoretical analysis and numerical simulations we shall identify here such interesting phenomena as
parametric resonance and a quasi-recurrence to a weakly excited starting state. To the best author's knowledge,
an idea about possibility of these effects in the system under consideration was not put forward previously by anyone.
Besides that, we will compare the results with other type parametric instabilities of a vortex ring, which were
found in recent studies to take place in harmonically trapped condensate with parabolic density profile
$\rho_{h}(z,r)=1-r^2-\alpha z^2$ near two definite values of the anisotropy parameter 
$\alpha^{(1)}= 8/5$ and $\alpha^{(2)} =16/7$.

\vspace{2mm}
{\bf Variational structure of equations}.
For our purposes it will be convenient to take angle $\varphi$ in the cylindrical coordinate system 
as the longitudinal parameter, while the two other coordinates will be considered as unknown functions  
$R(\varphi,t)$ and $Z(\varphi,t)$ (apparently, both $2\pi$-periodic on $\varphi$) which determine geometric shape
of the vortex ring at an arbitrary time moment. We restrict our study by axisymmetric density  profiles $\rho(z,r)$.
Equations of motion for $R(\varphi,t)$ and $Z(\varphi,t)$, equivalent to the vector equation (\ref{LIA}),
can be then written in a non-canonical Hamiltonian form,
\begin{eqnarray}
&&\rho(Z,R) R \dot Z=-\frac{\partial}{\partial\varphi}\frac{\rho(Z,R) R'}{\sqrt{R^2\!+\!R'^2\!+\!Z'^2}}\nonumber\\
&&\quad+\frac{\partial\rho(Z,R)}{\partial R} \sqrt{R^2\!+\!R'^2\!+\!Z'^2}
+\frac{\rho(Z,R) R}{\sqrt{R^2\!+\!R'^2\!+\!Z'^2}},
\label{dot_Z}
\end{eqnarray}
\begin{eqnarray}
&&-\rho(Z,R) R \dot R=-\frac{\partial}{\partial\varphi}\frac{\rho(Z,R) Z'}{\sqrt{R^2+R'^2+Z'^2}}\nonumber\\
&&\qquad+\frac{\partial\rho(Z,R)}{\partial Z} \sqrt{R^2+R'^2+Z'^2},
\label{dot_R}
\end{eqnarray}
where primes denote the partial derivatives on $\varphi$, and dots stand for time derivatives.
The corresponding Lagrangian has the following form 
\begin{equation}
{\cal L}=\int F(Z,R)\dot Z d\varphi -\int\rho(Z,R) \sqrt{R^2+R'^2+Z'^2}d\varphi,
\label{L_R_Z}
\end{equation}
where it is implied that for $\rho(z,r)=f(z,r^2/2)$ the function $F(Z,R)$  is determined by formula
\begin{equation}
F(Z,R)=\int_{U(z)}^{R^2/2} f(Z,u)du,
\end{equation}
and $U(z)$ can be chosen arbitrary. In particular, for  $r$-independent density profiles $\rho(z)$ we obtain 
$F=\rho(Z)R^2/2$,  while for condensate in a harmonic trap it is convenient to take  
$F=-(1-R^2-\alpha Z^2)^2/4=-\rho_h^2/4$.

\vspace{2mm}
{\bf Parametric instability}.
Let us first consider the case $\rho=\rho(z)$, when 
unperturbed propagation of a perfectly circular ring along  $z$ axis is described by solutions of the form
$R=R_0(t)$ and $Z=Z_0(t)$, which satisfy a simple system of ordinary differential equations
\begin{equation}
\dot Z_0=1/R_0, \qquad \dot R_0=-\rho'(Z_0)/\rho(Z_0). 
\end{equation}
Obviously, this system has integral of motion $R_0\rho(Z_0)=E=$ const.
Let us consider now the dynamics of small azimuthal deviations from the perfect shape, by writing
\begin{eqnarray}
R&=&R_0(t)+\sum_{m\ge 1}\Big[R_m e^{im\varphi}+R^*_m e^{-im\varphi}\Big],\\
Z&=&Z_0(t)+\sum_{m\ge 1}\Big[Z_m e^{im\varphi}+Z^*_m e^{-im\varphi}\Big],
\end{eqnarray}
where $R_m(t)$ and $Z_m(t)$ are small complex Fourier coefficients. A linearized system for them follows from
Eqs.(\ref{dot_Z})-(\ref{dot_R}). With taking into account the relation $d/dt=(1/R_0)d/dZ_0$ and the presence of
integral of motion  $R_0\rho(Z_0)=E$, we easily obtain
\begin{eqnarray}
\frac{d}{dZ_0}Z_m &=&\frac{\rho(Z_0)}{E}[m^2 -1]R_m,\\
-\frac{d}{dZ_0}R_m &=&\frac{\rho(Z_0)}{E}\Big[m^2 +\frac{E^2}{\rho^2(Z_0)}\Big(
\frac{\rho'(Z_0)}{\rho(Z_0)}\Big)'\Big]Z_m.
\end{eqnarray}
It is convenient to introduce here instead of $Z_0$ a new independent variable $\mu$ in accordance with 
$\rho(Z_0)dZ_0=d\mu$. Then the linearized system looks very simple:
\begin{eqnarray}
\frac{d Z_m}{d\mu} &=&\frac{1}{E}[m^2 -1]R_m,
\label{Z_mu}\\
-\frac{d R_m}{d\mu} &=&\frac{1}{E}\Big[m^2 +\frac{E^2}{f(\mu)}\frac{d^2 f(\mu)}{d\mu^2}\Big]Z_m,
\label{R_mu}
\end{eqnarray}
where function  $f(\mu)=\rho(Z_0(\mu))$ has been introduced. In our case this dependence is  $2\pi$-periodic,
so after reduction of (\ref{Z_mu})-(\ref{R_mu}) to a single differential equation of the second order we obtain 
a Hill equation,
\begin{equation}
\frac{d^2 Z_m}{d\mu^2}+\Big[\frac{m^2(m^2-1)}{E^2}+(m^2-1)\frac{f''(\mu)}{f(\mu)} \Big]Z_m=0,
\label{Hill}
\end{equation}
which is widely known as the main mathematical model describing parametric resonance in linear systems.
From here conditions for parametric resonance of order  $p=1,2,\dots$ immediately follow:
$E\approx E_m^{(p)}=2m\sqrt{m^2-1}/p$.
In this work we mainly concentrate on the case $m=2$, $p=1$. Let us note that at small values of the density
modulation depth $\epsilon\ll 1$ we have approximately ${f''(\mu)}/{f(\mu)}\approx \epsilon\cos \mu$, i.e.
the Hill equation takes form of the Mathieu equation. At the same time, the modulation depth of the nonuniform
coefficient in Eq.(\ref{Hill}) is equal to $12\epsilon$. The spatial increment of the instability at exact resonance
is given, as can be easily shown, by formula $\gamma^{(z)}\approx (3/2)\epsilon$. It corresponds to the growth
of the elliptic mode of the ring by a factor of $\exp(3\pi\epsilon)$ per one period of the density modulation.
Even with relatively small $\epsilon\sim 0.03$ we thus have a very rapid growth of deviations.

\vspace{2mm}
{\bf Numerical simulations}.
In order to investigate a nonlinear stage of the parametric instability development, solutions of the evolutionary
system (\ref{dot_Z})-(\ref{dot_R}) at $\rho=1-\epsilon \cos z$ with different initial conditions were found 
numerically by a pseudo-spectral method using a Runge-Kutta 4-th order procedure for the time stepping. 
Since it follows from the linear analysis of perturbations that at small $\epsilon$ near parametric resonance 
the dependencies $R_2(Z_0)$ and $Z_2(Z_0)$ have an oscillating character with period near $4\pi$, while their 
linear combinations $[R_2 -i(2/\sqrt{3}) Z_2]$ and  $[R^*_2 -i(2/\sqrt{3}) Z^*_2]$ are mainly proportional to 
$\exp(-i Z_0/2)$ (when higher harmonics are neglected), then for better understanding of the system dynamics it 
is useful to study behaviour of ``slow''complex-valued functions
\begin{eqnarray}
A_c&=&+2[\mbox{Re}(R_2) -i(2/\sqrt{3}) \mbox{Re}(Z_2)]\exp(i Z_0/2),\\
A_s&=&-2[\mbox{Im}(R_2) -i(2/\sqrt{3}) \mbox{Im}(Z_2)]\exp(i Z_0/2).
\end{eqnarray}
Let us note that $A_c$ and $A_s$ are complex envelopes for the amplitudes of standing modes $\cos 2\varphi$ and 
$\sin 2\varphi$ respectively, while $A_\pm=(A_c\mp iA_s)/2$ correspond to decomposition of elliptic perturbations
of the vortex ring on the propagating modes  $\exp(\pm 2i\varphi)$.

Two typical numerical examples of the ring perturbation dynamics are presented in Fig.1. The main features there
which catch our eye are the periodic synchronous returns of the system to a weakly excited state, alternating with
strongly deformed ring configurations, the last ones having different angular orientation on the $(x,y)$ plane.
Therefore, periodic is not each envelope taken separately, but their combination $\sqrt{|A_c|^2+|A_s|^2}$ 
which is independent on angle reading.
Only with increase of the parameter $\epsilon$ to values $\epsilon\sim 0.1$, the regular behaviour is destructed
(not shown in the figures). 

Such a recurrent dynamics is typical of autonomic integrable systems with a few degrees of freedom. 
Therefore it makes sense to derive a simplified model which could reproduce at least semi-quantitatively 
the dependencies observed in the numerical experiment.

\begin{figure}
\begin{center}
\epsfig{file=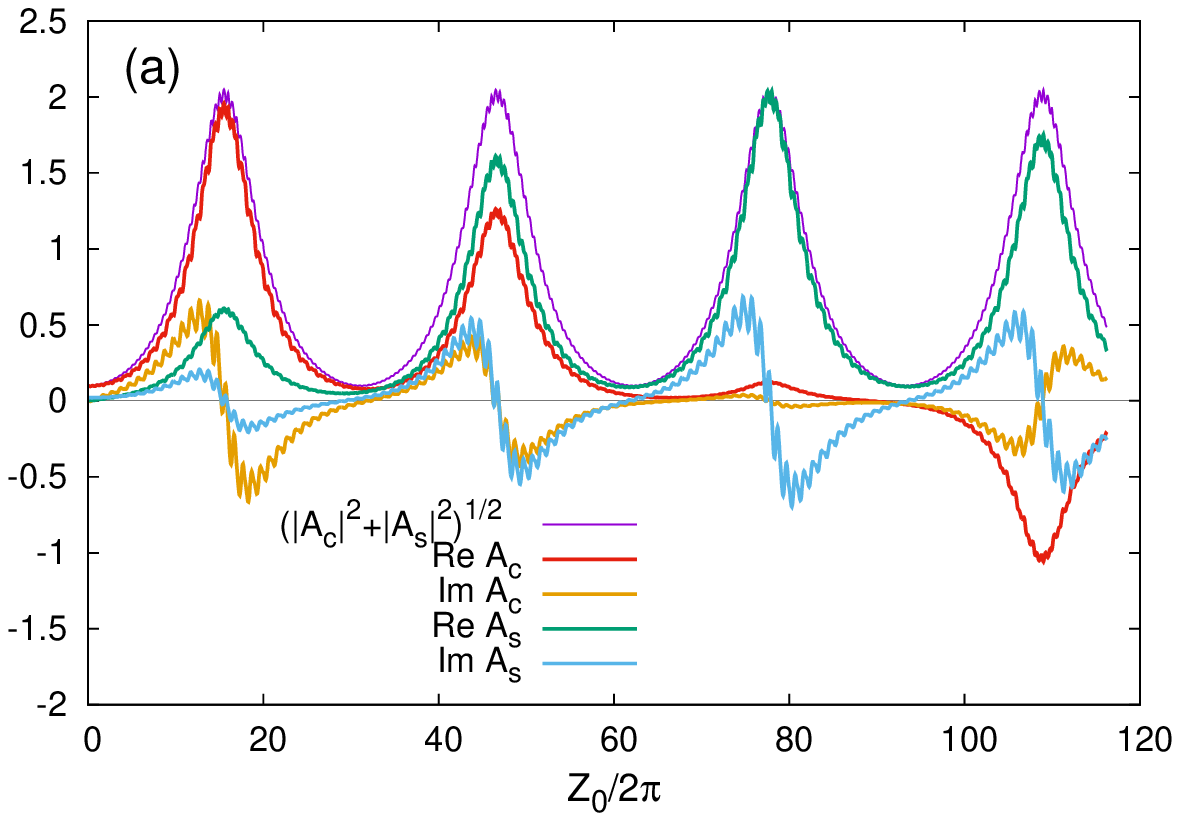, width=80mm}\\
\epsfig{file=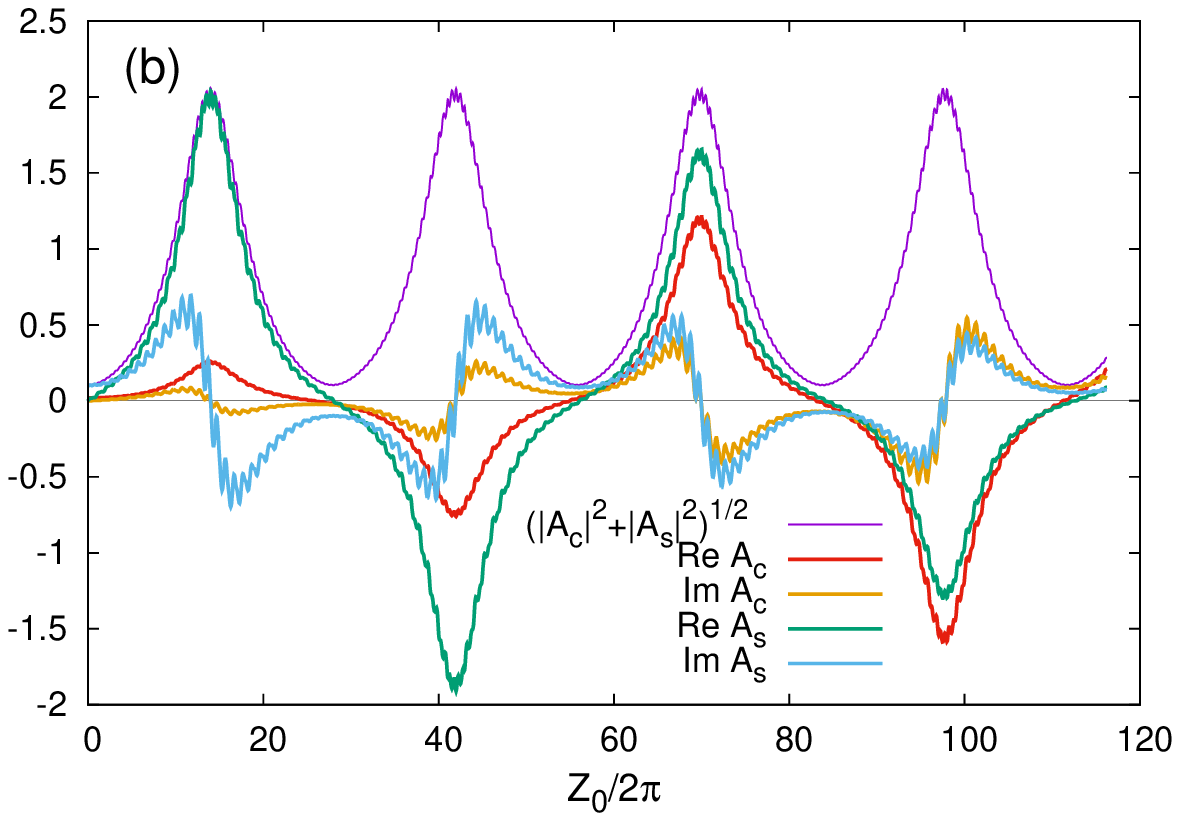, width=80mm}
\end{center}
\caption{Two examples of the vortex ring parametric instability development and its return to a weakly excited state
at $\epsilon=0.03$. The residual oscillations on the curves are caused by the presence of higher harmonics,
which are typical for parametrically unstable systems. Initial conditions in  case (a): 
$R(0)=4\sqrt{3} +0.10\cos(2\varphi)$, $Z(0)= -0.02(\sqrt{3}/2)\sin(2\varphi)$, 
it corresponds to $A_c(0)=0.10$, $A_s(0)=0.02i$; in case (b): 
$R(0)=4\sqrt{3} +0.02\cos(2\varphi)$, $Z(0)= -0.10(\sqrt{3}/2)\sin(2\varphi)$, 
it corresponds to $A_c(0)=0.02$, $A_s(0)=0.10i$. }
\label{ReA_ImA_Z0} 
\end{figure}

\vspace{2mm}
{\bf Explanation of the recurrence phenomenon}.
In order to explain theoretically the recurrent dynamics of ring deformations, 
we introduce new canonically conjugate variables
\begin{eqnarray}
S&=&\frac{R^2}{2}\rho^2(Z)=S_0+\sum_{m\ge 1}[ S_m e^{im\varphi}+S_m^*e^{-im\varphi}], \\
\chi&=&\int_0^Z \frac{dz}{\rho(z)} =\chi_0+\sum_{m\ge 1} [\chi_m e^{im\varphi}+\chi_m^*e^{-im\varphi}],
\end{eqnarray}
and expand on small disturbances the corresponding Hamiltonian of the ring,
\begin{equation}
H=\int\sqrt{2S+g^4(\chi)\chi'^2 +g^2(\chi)\Big(\frac{\sqrt{2S}}{g(\chi)}\Big)'^2}d\varphi,
\end{equation}
with $g(\chi(z))=\rho(z)$ [at small $\epsilon$ we have $g(\chi)=1-\epsilon\cos(\chi)+{\cal O}(\epsilon^2)$].
At that we obtain $H/(2\pi)=H^{\{0\}}+\sum_{m\ge 1} H_m^{\{2\}} +H^{\{3\}} +H^{\{4\}} \dots$, where $H^{(0)}=\sqrt{2S_0}$,
\begin{eqnarray}
&&H_m^{\{2\}}=-\frac{|S_m|^2}{\sqrt{(2S_0)^3}} +\frac{m^2}{\sqrt{2S_0}}\Bigg[g_0^4|\chi_m|^2\nonumber\\
&&+\frac{|S_m|^2}{2S_0}
-\frac{g_0'}{g_0}(S_m\chi_m^* +S_m^*\chi_m)+\frac{2S_0g_0'^2}{g_0^2} |\chi_m|^2\Bigg],
\end{eqnarray}
and $g_0=g(\chi_0)$. Let us separate in $H_m^{\{2\}}$ the terms of zeroth order on $\epsilon$
and in a standard way construct on them the normal complex variables
\begin{equation}
a_m=\frac{\sqrt{m^2-1}(2S_0)^{-3/4} S_m-im(2S_0)^{-1/4}\chi_m}{\sqrt{2\omega_m}},
\end{equation}
\begin{equation}
a_{-m}=\frac{\sqrt{m^2-1}(2S_0)^{-3/4} S_m^*-im(2S_0)^{-1/4}\chi_m^*}{\sqrt{2\omega_m}},
\end{equation}
where the frequency in spatially uniform system is
\begin{equation}
\omega_m=\frac{m\sqrt{m^2 -1}}{2S_0}.
\end{equation}
The variable $\chi_0$ is slightly renormalized at that, but we keep the same notation.

It is very important that at $\epsilon=0$ our system is completely integrable. Therefore there exist such
renormalized normal variables $b_m =a_m+{\cal O}\{a^2\}$, that three-wave interactions are excluded, while
fourth-order terms have the form $\tilde H^{\{4\}}=(1/2)\sum_{k,n}W_{kn}|b_k|^2 |b_n|^2$. 

Let us consider mode excitations for resonance number  $m$ at $p=1$ and introduce slow envelopes for 
the corresponding normal variables:
\begin{equation}
b_m=b_+\exp(-i\chi_0/2), \quad b_{-m}=b_-\exp(-i\chi_0/2).
\label{a_pm}
\end{equation}
After that we average  $\tilde H_m^{\{2\}}$ on the density oscillations with an accuracy up to
the first order on  $\epsilon$. Nontrivial averaging is required for the term proportional to 
$-4\epsilon \cos(\chi_0)|\chi_m|^2$, and also for  $-\epsilon \sin(\chi_0)(S_m\chi^*_m+S_m^*\chi_m)$. 
As the result of substitution (\ref{a_pm}) and subsequent averaging, an effective Lagrangian takes the following form:
\begin{eqnarray}
L&\approx& S_0 \dot \chi_0 +i\dot b_+ b_+^* +i\dot b_- b_-^* 
+\frac{\dot\chi_0}{2}(|b_+|^2 + |b_-|^2)\nonumber\\
&&-\sqrt{2S_0}-\frac{m\sqrt{m^2-1}}{2S_0}(|b_+|^2 + |b_-|^2)\nonumber\\
&&+\epsilon\frac{\sqrt{m^2-1}}{4m}(b_+ b_- + b_+^* b_-^*) \nonumber\\
&&-T(|b_+|^4+ |b_-|^4) - W|b_+|^2 |b_-|^2 .
\end{eqnarray}
It is important that we have here an integrable Hamiltonian system with three degrees of freedom.
Apparent integrals of motion, besides the Hamiltonian itself, are 
\begin{eqnarray}
S_0+\frac{1}{2}(|b_+|^2 + |b_-|^2)=I&=&\mbox{const},
\label{I}\\
|b_+|^2 - |b_-|^2=D&=&\mbox{const}.
\label{D}
\end{eqnarray}
Formula (\ref{I}) shows that a mean size of the ring is decreased as its deformation is increased.
Of course, it is consistent with conservation of the total energy.
The conservation law (\ref{D}) is actually for $z$-component of the angular momentum.
Excluding  $S_0$, we obtain an effective Hamiltonian of perturbations in the form
\begin{eqnarray}
&&\tilde H= \frac{m\sqrt{m^2-1}(|b_+|^2 + |b_-|^2)}{(2I-|b_+|^2 - |b_-|^2)}+\sqrt{2I-|b_+|^2 - |b_-|^2}\nonumber\\
&&\qquad-\epsilon\frac{\sqrt{m^2-1}}{4m}(b_+ b_- + b_+^* b_-^*) \nonumber\\
&&\qquad+T(|b_+|^4+ |b_-|^4) + W|b_+|^2 |b_-|^2 .
\label{Hpm}
\end{eqnarray}
In terms of canonically conjugate variables $N=|b_+|^2 + |b_-|^2$ and $\Phi=[\arg(b_+)+\arg(b_-)]/2$ it is reduced to 
\begin{eqnarray}
&&\tilde H= \frac{m\sqrt{m^2-1}N}{(2I-N)} -\epsilon\frac{\sqrt{m^2-1}}{4m}\sqrt{N^2-D^2}\cos(2\Phi) 
\nonumber\\
&&\qquad+\sqrt{2I-N}+\frac{T}{2}(N^2+ D^2) + \frac{W}{4}(N^2-D^2).
\end{eqnarray}
The recurrence phenomenon then corresponds to quasi-closed phase trajectories in the complex plane of variable
$C=\sqrt{(|A_+|^2+|A_-|^2)/2 }(A_+A_-)/|A_+A_-|$, as shown in Fig.2. Of course, the real phase trajectory
only ``in average'' is described by the simplified model.

\begin{figure}
\begin{center}
\epsfig{file=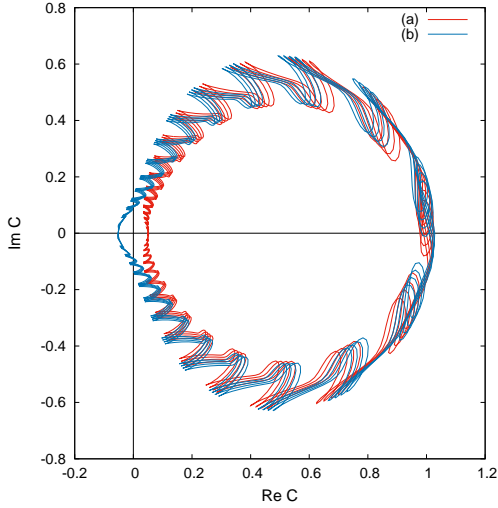, width=70mm}
\end{center}
\caption{Quasi-closed phase trajectories in plane $C$, corresponding to two numerical experiments presented in
Fig.1.}
\label{ReC_ImC} 
\end{figure}

It is necessary also to say that applicability of the Hamiltonian (\ref{Hpm}) is limited by close-to-resonance
values of the parameter  $I$. Taking $2I=4m^2(m^2-1)+\delta$  and making expansion up to 4-th order on $b_\pm$,
we obtain a quite simplified model Hamiltonian as written in Abstract. Its quadratic part corresponds to instability 
near the resonance.

\vspace{2mm}

\begin{figure}
\begin{center}
\epsfig{file=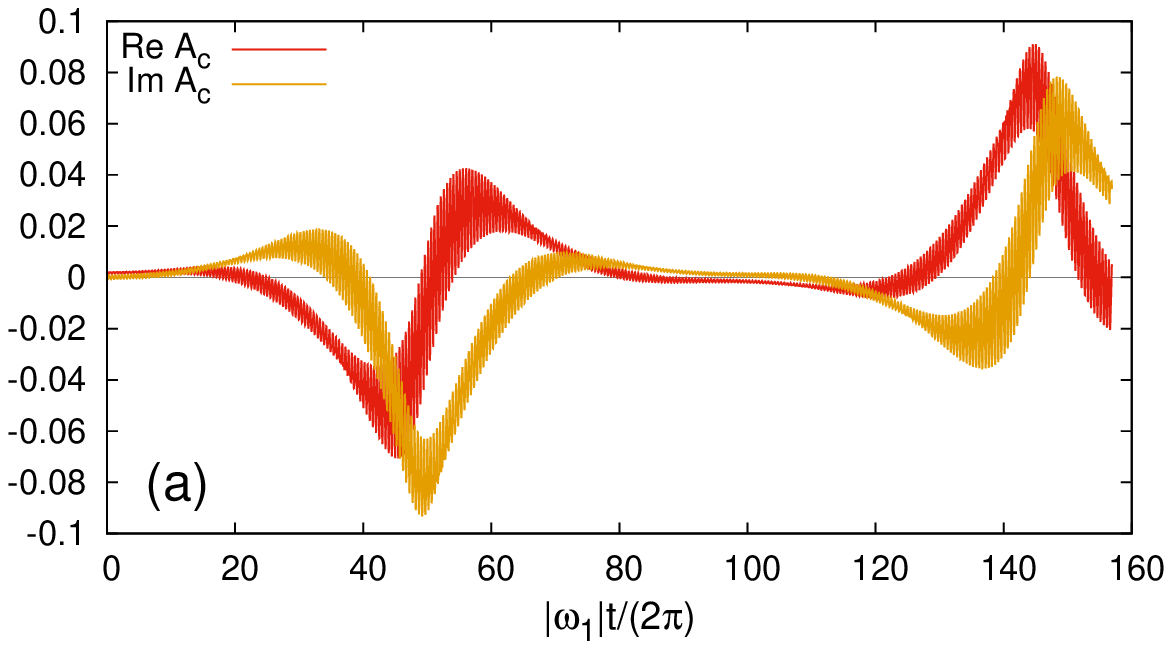, width=75mm}\\
\epsfig{file=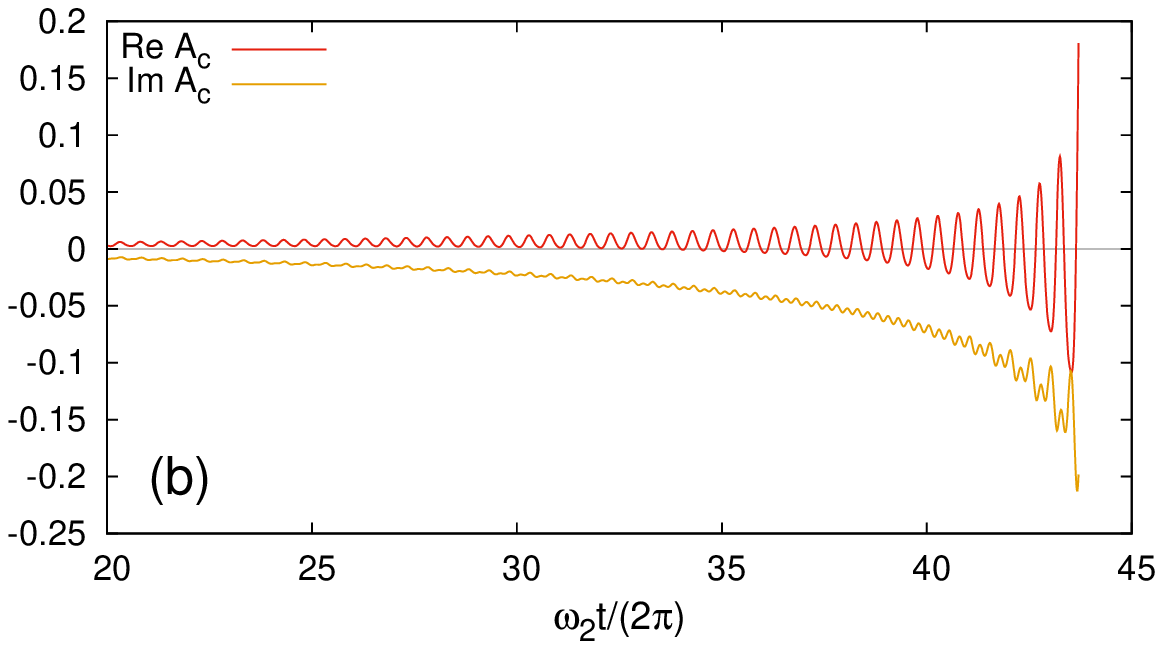, width=75mm}
\end{center}
\caption{Two types of nonlinear stage of parametric instabilities of a vortex ring in condensate with density
$\rho=1-r^2-\alpha z^2$, observed in numerical experiments. Shown are properly defined envelopes of unstable modes.
The oscillations on the curves are caused by higher harmonics. In the first case $\alpha =8/5$, 
$R(0)=0.88/\sqrt{3} + 0.002\cos(\varphi)$, $Z(0)=0$. In the second case  $\alpha =16/7$, 
$R(0)=0.95/\sqrt{3} + 0.002\cos(2\varphi)$, $Z(0)=0$.}
\label{trap} 
\end{figure}

{\bf Comparison to vortex ring in a harmonic trap}. Since we speak here about parametric instabilities of 
a quantum vortex ring, it makes sense to compare the above described instability mechanism to a situation where
the condensate is bounded in space, and a finite motion of perfect circular ring is described by two periodic
functions $R_0(t)$ and $Z_0(t)$. Specifically, we consider a density profile $\rho_{h}(z,r)=1-r^2-\alpha z^2$
which is typical for harmonically trapped  Bose-Einstein condensates in the Thomas-Fermi regime.
Each phase trajectory of the perfect ring encompasses point $R_{\rm st}=1/\sqrt{3}$, $Z_{\rm st}=0$, while the squared 
eigenfrequencies of small oscillations are given by expression first obtained in Ref. \cite{ring_istability}:
\begin{equation}
\omega_m^2=9(m^2-3)(m^2-\alpha).
\label{ww_trap}
\end{equation}
It follows from here that in the range $1<\alpha<4$ all the modes are linearly stable. In our approach this result
is easily reproduced by introducing canonical variables  $Q=1-R^2-\alpha Z^2$ and $P=QZ/2$. The Hamiltonian of perfect
ring is defined then by rather elegant expression,
\begin{equation}
{\cal H}_0=\sqrt{Q_0^2-4\alpha P_0^2-Q_0^3},
\end{equation}
and the stationary point is $Q_{\rm st}=2/3$, $P_{\rm st}=0$. Let us study small deviations of  ring from the equilibrium,
by writing the local induction Hamiltonian $(2\pi)^{-1}\int\rho(Z,R)\sqrt{(R^2+R'^2+Z'^2)}d\varphi$ in terms of the new
variables and expanding on powers of small functions  $q=Q(\varphi,t)-2/3$ and $p=P(\varphi,t)$. We thus obtain
\begin{eqnarray}
&&{\cal H}^{\{2\}}\!=\!\sqrt{3}\sum_{m}\!\Big[\frac{1}{4}(m^2\!-\!3)|q_m|^2\!+\!3(m^2\!-\!\alpha)|p_m|^2\Big],\\
&&{\cal H}^{\{3\}}=-\frac{1}{2\pi}\int\Big[\frac{3\sqrt{3}}{4}q^3 +\frac{9\sqrt{3}}{2}(\alpha-1)p^2q''\Big]d\varphi.
\end{eqnarray}
The term ${\cal H}^{\{2\}}$ gives formula (\ref{ww_trap}). In terms of normal complex variables defined as
\begin{equation}
a_m=\frac{(\sqrt{3}|m^2\!-\!3|/2)^{\frac{1}{2}}q_m+i(6\sqrt{3}|m^2-\alpha|)^{\frac{1}{2}}p_m}{\sqrt{2|\omega_m|}},
\end{equation}
the quadratic Hamiltonian  ${\cal H}^{\{2\}}=\sum_{m}\omega_m|a_m|^2$. It is very important that the two first frequencies
$\omega_0$ and $\omega_1$ have the negative sign, while at  $|m|\ge 2$ all $\omega_{m}$ are positive (which fact was not
taken into account by the authors of Ref.\cite{ring_istability}, since they did not use the Hamiltonian method).
Therefore at definite values of the anisotropy parameter $\alpha$, nonlinear resonances arise between some modes, 
leading to parametric instabilities. In particular, the condition $\omega_0\approx 2\omega_1$ takes place
near  $\alpha^{(1)}=8/5$, and then nonlinear resonance processes occur which are described by interaction 
of the form $V^{(1)}(a_0^*a_1 a_{-1} + a_0 a_1^* a_{-1}^*)$, while at $\alpha\approx\alpha^{(2)}=16/7$, when
$\omega_0\approx -2\omega_2$, in resonance are the processes corresponding to interaction in the form
$V^{(2)}(a_0 a_2 a_{-2} + a_0^* a_2^* a_{-2}^*)$. In both situations, a weakly nonlinear dynamics is approximately 
described by integrable Hamiltonians of standard form:
\begin{eqnarray}
{\cal H}^{(1)}&=&(\delta^{(1)}-2\Omega^{(1)})|a_0|^2-\Omega^{(1)}(|a_1|^2+|a_{-1}|^2)\nonumber\\
&&+V^{(1)}(a_0^*a_1 a_{-1} + a_0 a_1^* a_{-1}^*),
\end{eqnarray}
\begin{eqnarray}
{\cal H}^{(2)}&=&(\delta^{(2)}-2\Omega^{(2)})|a_0|^2+\Omega^{(2)}(|a_2|^2+|a_{-2}|^2)\nonumber\\
&&+V^{(2)}(a_0 a_2 a_{-2} + a_0^* a_2^* a_{-2}^*),
\end{eqnarray}
where $\delta^{(1)}$ and $\delta^{(2)}$ are small frequency detuning parameters, and the coefficients  
$V^{(1)}$, $V^{(2)}$ can be calculated by re-writing ${\cal H}^{\{3\}}$ in terms of  $a_m$.
In the first case there are additional integrals of motion  $|a_0|^2+|a_1|^2=s_+$ and $|a_0|^2+|a_{-1}|^2=s_-$,
so the system remains in a weakly nonlinear regime, and a periodic recurrence of ring to an almost 
axisymmetric state occurs, as it is shown in Fig.3a. In the second case, the additional integrals of motion  are
$|a_0|^2-|a_2|^2=d_+$ and $|a_0|^2-|a_{-2}|^2=d_-$, and the parametric instability has an explosive character,
as it is seen in Fig.3b. In fact, as numerical simulations of system (\ref{dot_Z})-(\ref{dot_R}) at  
$\rho=1-r^2-\alpha z^2$ demonstrate, at the final stage of the explosive instability so large deformation of 
the vortex ring is attained that some its parts closely approach the Thomas-Fermi surface (which is an effective
condensate boundary), and there the hydrodynamic anelastic approximation certainly fails.

\vspace{2mm}
{\bf Conclusions}.
Thus, in this work for the first time we have predicted the parametric instability of oscillations of quantum vortex 
ring in a spatially periodic Bose-Einstein condensate at definite sizes of the ring, and also in harmonically
trapped condensate --- at definite values ot the trap anisotropy. In all the cases, we numerically simulated 
nonlinear stages of the instabilities. The found here phenomenon of quasi-recurrence was theoretically explained. 
Such kind nontrivial behaviour of vortex ring definitely deserves further studies within 
more accurate models. In particular, it is very desirable to reproduce the parametric instability at moderate 
values of $\Lambda$ immediately in a numerical solution of the 3D Gross-Pitaevskii equation with periodic 
external potential, as well as in anisotropic harmonic potential.
After that, organization of some real-world experiments could become actual.

\end{document}